# Main-Group Metal Elements as Promising Active Centers for Single-Atom Catalysts

Qian Wu[1], Baibiao Huang[1], Ying Dai[1*], Thomas Heine[2,3,4*], and Yandong Ma[1*]

[1]School of Physics, State Key Laboratory of Crystal Materials, Shandong University, Shandanan Str. 27, Jinan 250100, China

[2]TU Dresden Fakultät für Chemie und Lebensmittelchemie, Bergstraße 66c, 01062 Dresden, Germany

[3]Helmholtz-Zentrum Dresden-Rossendorf, Forschungsstelle Leipzig Permoserstraße 15, 04318 Leipzig, Germany

[4]Department of Chemistry, Yonsei University, Seodaemun-gu, Seoul 120-749, Republic of Korea

E-mail: daiy60@sina.com (Y.D.); thomas.heine@tu-dresden.de (T.H.);

yandong.ma@sdu.edu.cn (Y.M.)

**Abstract**

Current research efforts on single-atom catalysts (SACs) exclusively focus on nonmetal or transition-metal atoms as active centers, while employing main-group metal elements is seemingly excluded because their delocalized s/p-bands are prone to yield a broadened resonance for the interaction with adsorbates. Here, we use high-throughput first-principles calculations to investigate the possible incorporation of Mg, Al and Ga to form graphene-based SACs for NO reduction reaction (NORR) toward $NH_3$. 51 SAC candidates with different metal coordination environment have been computationally screened employing a rationally designed four-step process, yielding six SACs with high catalytic activity and NORR selectivity. The performance is rationalized by the modulation of s/p-band filling of the main-group metals. The adsorption free energy of NO is identified as efficient descriptor for such SACs. The underlying physical mechanism is revealed and generally applicable to other main group metal SACs. These fundamental insights extend SACs to main-group metal elements.

Table of content

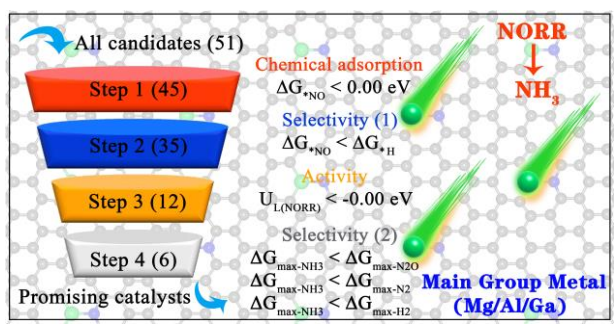

## 1. Introduction

Single-atom catalysts (SACs) have drawn increasingly intense attention in electrocatalysis, as they have great potential to provide unmatched high activity and selectivity for multiple chemical reactions.[1-3] The catalytic activity of SAC active centers is strongly correlated to their coordination environment.[4-7] In recent years, much progress has been made, and various SACs have been developed both theoretically [8-11] and experimentally.[12-15] The catalytic properties of some theoretically predicted SACs have been verified experimentally.[16-18] Nonetheless, these investigations mostly focus on nonmetal or transition-metal atoms as SAC active centers.[19] While the catalytic activity of the nonmetal SACs is caused by hybridized s-p orbitals, transition metal SACs benefit from the localized character of d orbitals.[20-23] The introduction of elements besides nonmetal and transition metal atoms as SAC active centers remains to be explored.

In contrast to transition metals, main-group metal elements such as Mg, Al and Ga, generally exhibit a delocalized s/p-band, which is prone to broaden the adsorbate states.[22,24,25] Such interaction is thought to yield either too strong or too weak adsorption for adsorbates. This would poison active centers or fail to activate adsorbates, which thus seemingly excludes any possibility to achieve highly performing SACs based on main-group metal elements.[25] Nevertheless, recently main-group metal elements have been employed as co-electrocatalysts in dual atomic catalysis, where they act as promising separator and facilitate the electroactivity of the connected transition metal atoms.[19] Moreover, Mg ion that is contained in enzymes has been proven to show suitable affinity for oxygenated species.[26-29] These observations suggest that, despite the unbefitting delocalized s/p-band, main-group metal elements might still offer the potential to be active centers in SACs.

Among various electrocatalysis reactions, the direct electrochemical NO reduction reaction (NORR) toward $NH_3$ receives special attention as it combines both NO removal and $NH_3$ synthesis.[30-33] For NO removal, the conventional technology is based on selective catalytic reduction, which consumes valuable $NH_3$ or $H_2$.[34,35] The alternative, electrochemical $NH_3$ synthesis from $N_2$, suffers from low reaction rate and Faradaic efficiency.[36-39] Undoubtedly, a successful realization of NORR, converting NO into NH$_3$ directly, will be a breakthrough.[20,30,40] Though highly valuable, the direct NO-to-NH$_3$ conversion is essentially unexplored, and only a few NORR electrocatalysts have been reported to date.[30-33,40,41] Exploring suitable NORR catalysts thus is of great economic interest and scientific importance, but a formidable task.

In this work, we systematically investigate the possibility of introducing main group metal elements (i.e., Mg, Al, and Ga) to form graphene-based SACs and investigate their suitability to drive the NORR. In strong contrast to expectations, i.e. that suitable active centers are either nonmetal and transition-metal atoms, we demonstrate that indeed main-group metal elements can serve as promising active centers for SACs toward NORR. Through high-throughput screening based on first-principles calculations, six SAC systems out of 51 candidates have been selected as they exhibit superior catalytic activity and selectivity for NORR toward NH$_3$. Importantly, the NORR process can occur spontaneously in these systems, guaranteeing fast kinetics. The excellent performances of these systems are closely related to the modulation of s/p-band filling of the main-group metal centers derived from regulating coordination environment. A rational and generally applicable four-step screening principle is mapped out. Furthermore, we discover that the adsorption free energy of NO is an efficient catalytic descriptor for such SACs. Our findings provide an effective guidance for understanding the potential introduction of main-group metal elements in SACs, and shed light on further development of NORR catalysts.

## 2. Results and Discussion

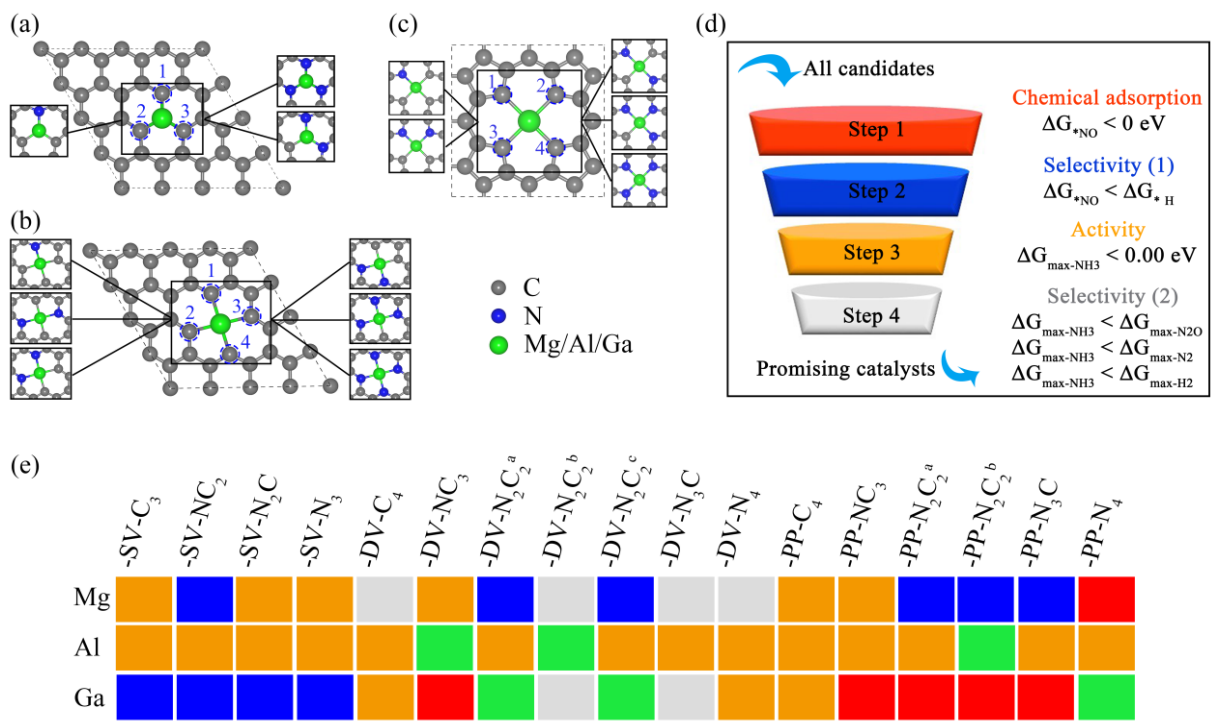

**Figure 1.** Schematic illustrations of single Mg/Al/Ga atom supported on N-doped graphene with different coordination environments and a screening strategy flowchart for NORR toward NH$_3$. a)

Single vacancy site coordinated with three C/N atoms (SV-$N_{3-n}C_n$, n=0~3). b) Double vacancy site coordinated with four C/N atoms (DV-$N_{4-n}C_n$, n=0~4). c) Porphyrin site coordinated with four C/N atoms (PP-$N_{4-n}C_n$, n=0~4). The unit cells are marked in dotted lines, and the solid rectangles highlight the local configurations with different numbers of N atoms. Atoms marked with dotted circles in (a-c) indicate the replaceable C atoms. d) Flowchart of the screening scheme to identify the promising SACs with main group metal active centers for NORR toward $NH_3$. (e) Color code map corresponding to the flowchart of the screening scheme shown in (d). The red, blue, orange and gray color codes indicate the eliminated systems from the first, second, third and fourth screening steps, respectively, and the green color code indicates the final candidates.

To support the dispersed Mg/Al/Ga atoms, we have chosen N-doped graphene as substrate in light of the following reasons. First, N-doped graphene is well defined and has been demonstrated to act as excellent substrate for different SACs both theoretically and experimentally. It merits structural flexibility and thermostability, so it can be used in harsh chemical environments.[4,5,42-46] Second, it has different coordination modes and thus offers various coordination environments for active centers. Third, N-doped graphene exhibits excellent conductivity, facilitating fast electron transfer during electrocatalysis. **Figure 1**a-c illustrates the schematic configurations of single Mg/Al/Ga atom supported on graphene with the coordination of N atoms. By regulating the N-coordination, 17 configurations are constructed for each case, including four SV-$N_{3-n}C_n$, seven DV-$N_{4-n}C_n$ and six PP-$N_{4-n}C_n$ (Figure 1a-c), which results in a total of 51 SAC candidates. The corresponding optimized crystal structures are shown in Figures S1-S3 of the Supporting Information. The Bader charge analysis indicates that the supported main group atom donates electrons to the substrate for all these systems, as listed in Table S1. The resultant positively charged metal atoms act as Lewis acids and hence become active sites for catalytic reactions.[47-49]

The electrochemical NORR for $NH_3$ synthesis is a complex process involving five hydrogenation steps, i.e., NO + 5H$^+$ + 5e$^-$ = $NH_3$ + $H_2O$. To identify suitable NORR SACs, we design a four-step screening strategy, as schematically illustrated in Figure 1d. First of all, the NO molecule should be chemisorbed and activated on the SAC center ($\Delta G_{*NO} < 0$, where $\Delta G_{*NO}$ is the Gibbs free energy of activated *NO adsorption), which is required to facilitate the subsequent hydrogenation process. Second, the competitive proton adsorption should be suppressed to protect the electrocatalyst from being poisoned ($\Delta G_{*NO} < \Delta G_{*H}$, where $\Delta G_{*H}$ is the Gibbs free energy change for H adsorption). Third, the criterion of $\Delta G_{max-NH3} < 0$ ($\Delta G_{max-NH3}$ is the maximum Gibbs free energy change of hydrogenation steps for $NH_3$ synthesis via the most favorable reaction mechanism) is required for fast kinetics. Finally, $G_{max-NH3} < \Delta G_{max-N2O}$, $\Delta G_{max-NH3} < \Delta G_{max-N2}$ and $\Delta G_{max-NH3} < \Delta G_{max-H2}$ ($\Delta G_{max-N2O/N2/H2}$ is the maximum Gibbs free energy change of hydrogenation steps for $N_2O/N_2/H_2$ synthesis via the corresponding most favorable reaction mechanism) are adopted to select systems of high $NH_3$ selectivity. Following this four-step screening strategy, the highly efficient NORR SACs with main-group metal elements as active centers for $NH_3$ synthesis can be identified using first-principles calculations.

For NORR process, the first reaction step is adsorption (and activation) of NO, which plays a critical role for the subsequent reactions. Given the non-equivalence between N and O atoms, three possible adsorption patterns of NO are considered, i.e., N-end, O-end and NO-side patterns (**Figure 2**a). This gives rise to 153 adsorption configurations in total. The corresponding $\Delta G_{*NO}$ for all these configurations are shown in Figure 2b and Tables S2-S4. Intriguingly, the N-end pattern is most preferred in energy for almost all SAC candidates, except for Mg-SV-$C_3$ (O-end) and for Mg-PP-$C_4$/-N$C_3$, Al-PP-$C_4$ and Ga-DV-$C_4$ (NO-side), see Figures S4-S6. This behavior is expected due to the smaller electronegativity of N atom compared to O atom, making N atom more attractive to coordinate to the SAC Lewis acid site. Note that to guarantee sufficient activation of the adsorbed NO, spontaneous chemisorption of NO ($\Delta G_{*NO} < 0$) is regarded as a necessary prerequisite for NORR process. Thus, following the first screening stage, the systems shown in red color in Figure 1e are discarded ($\Delta G_{*NO} > 0$). The other SAC candidates show negative adsorption free energies between −0.08 and −3.21 eV.

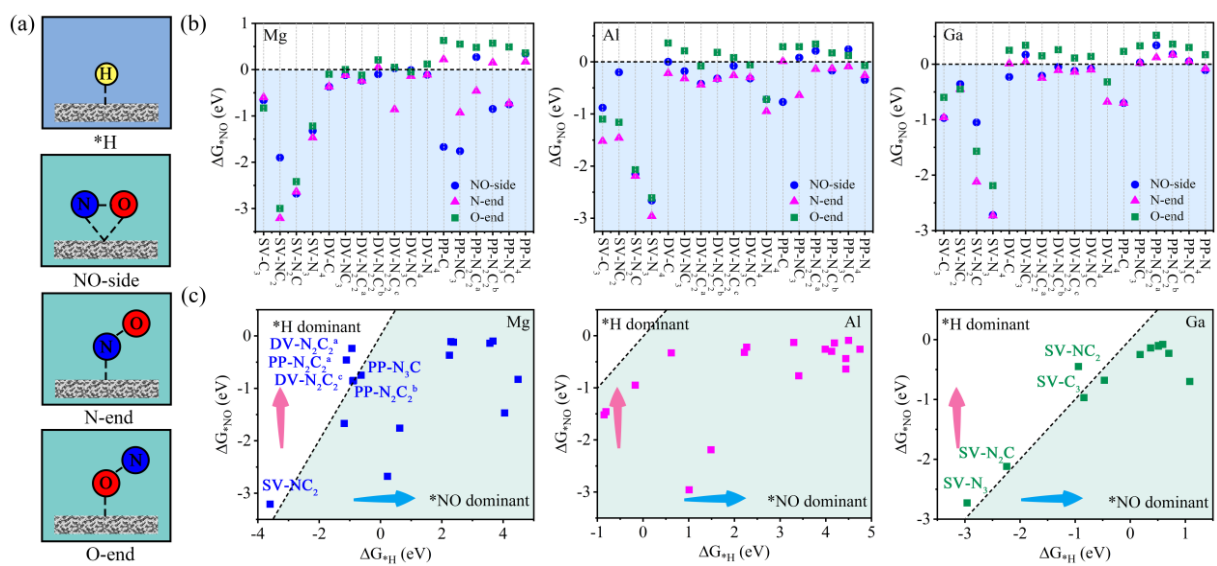

**Figure 2.** Adsorption performance of NO molecule and selectivity between the adsorptions of NO and H on SAC candidates. a) The possible adsorption patterns of H and NO (i.e., N-end, O-end and NO-side patterns). b) Gibbs free energy changes for NO adsorption with N-end, O-end and NO-side patterns. c) Comparison of the adsorption free energies of H proton and NO molecule with the corresponding most favorable adsorption pattern.

To ensure an effective chemisorption of NO, the competitive adsorption of proton should be excluded as well, which we consider by calculating the selectivity between the adsorptions of both species on the remaining SAC candidates. The Gibbs free energy changes for proton adsorption are summarized in Table S5, and the comparison of adsorption free energies between NO and H is presented in Figure 2c. We rule out all systems where protons are preferentially or similarly strongly adsorbed, as they would refer to proton-poisoned SAC (blue colored in Figure 1e).

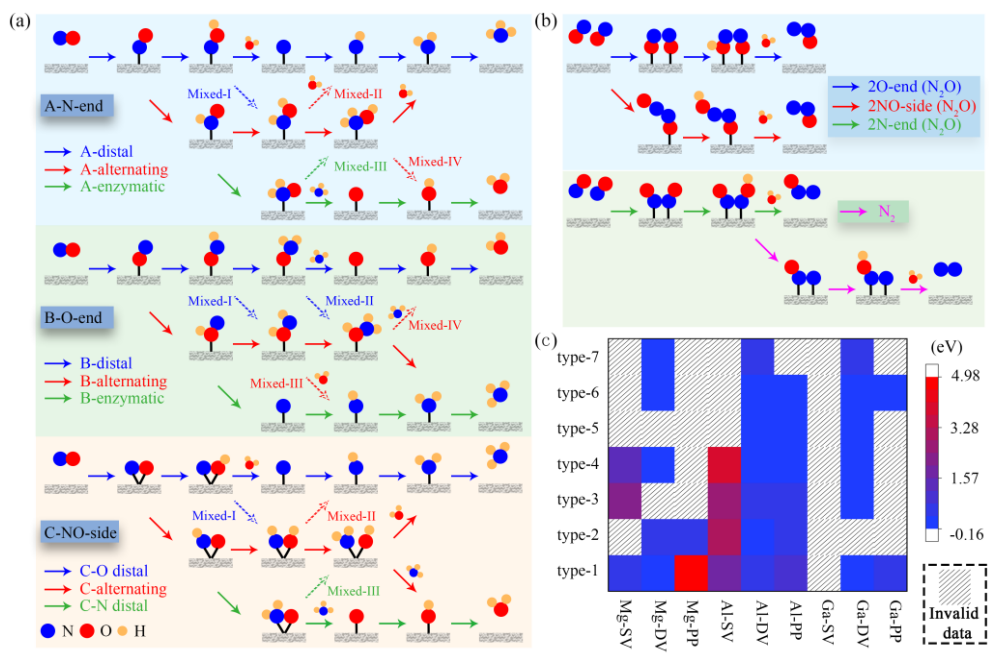

**Figure 3.** Reaction mechanism and electrocatalytic activity of NORR on SAC candidates. a) Schematic depiction of possible NORR pathways toward $NH_3$ synthesis. b) Schematic depiction of possible NORR pathways toward $N_2O$ and $N_2$ synthesis. c) $\Delta G_{max\text{-}NH_3}$ on the remaining 35 SAC candidates via the corresponding most favorable reaction pathways.

Then, we investigate the subsequent hydrogenation steps of NORR process toward $NH_3$ synthesis for the remaining 35 SAC candidates. Depending on adsorption patterns of NO, the reaction pathways of NORR toward $NH_3$ can follow the A/B-distal, A/B/C-alternating, A/B-enzymatic and C-O/N distal mechanisms (**Figure 3**a). Remember that the NO adsorption patterns are O-end for Mg-SV-$C_3$ and NO-side for Mg-PP-$C_4$/$NC_3$, Al-PP-$C_4$ and Ga-DV-$C_4$, the B-distal/alternating/enzymatic and C-O distal/alternating/N-distal mechanisms are considered for them, respectively. While for the other systems, as the adsorption pattern is N-end, the A-distal/alternating/enzymatic mechanisms are explored. Also, the corresponding mixed mechanisms shown in Figure 3a are investigated for all these systems. To determine the most favorable reaction mechanism for the NORR process, we comprehensively calculate the Gibbs free energies of all the possible reaction intermediates. And we plot the $\Delta G_{max\text{-}NH_3}$ in Figure 3c. Here, to ensure fast kinetics, we set $\Delta G_{max\text{-}NH_3} < 0$ as strict screening criterion, so the reaction can proceed spontaneously. As shown in Figure 3c, the systems depicted in orange in Figure 1e are eliminated. We wish to emphasize that although Mg-DV-$NC_3$, Mg-PP-$NC_3$,

Al-DV-C$_4$/-N$_2$C$_2$$^a$/-N$_2$C$_2$$^c$/-DV-N$_3$C/-N$_4$, Al-PP-N$_2$C$_2$$^a$/-N$_3$C/N$_4$, and Ga-DV-C$_4$ are excluded here, they exhibit low limiting-potentials with 0.0 < $\Delta G_{max-NH3}$ < 0.5 eV. For the remaining 12 SAC candidates (shown in gray and green colors in Figure 1e), the NORR process occurs spontaneously. It should be noted that these candidates with high activity are all based on DV-N$_{4-n}$C$_n$ or PP-N$_{4-n}$C$_n$ cases, while SV-N$_{3-n}$C$_n$ is completely excluded.

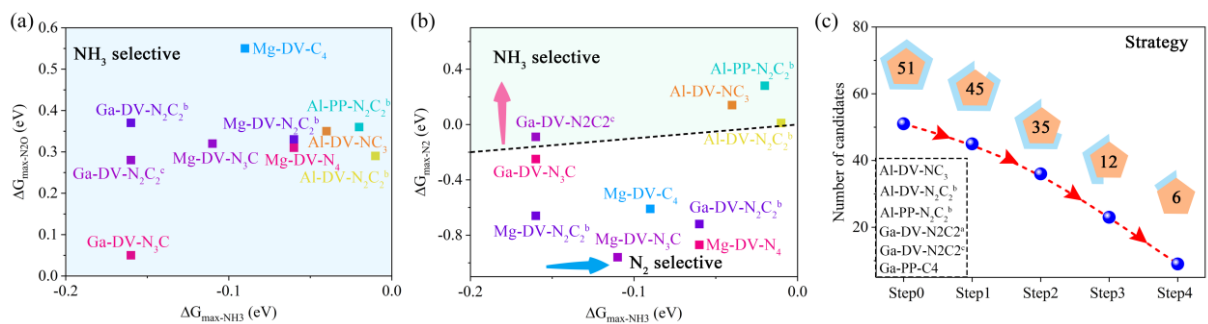

**Figure 4.** Electrocatalytic selectivity and summary of the screening strategy for NORR toward NH$_3$. a) $\Delta G_{max-NH3}$ vs $\Delta G_{max-N2O}$. b) $\Delta G_{max-NH3}$ vs $\Delta G_{max-N2}$. c) Illustration of the screening strategy applied to identify high-efficient electrocatalysts for NORR. Insets in (c) shows the corresponding selected high-efficient SAC catalysts.

To inspect the selectivity of NORR toward NH$_3$ synthesis for the remaining 12 SAC candidates, we further investigate the possible formation of byproducts N$_2$O and N$_2$ during the NORR process. The reaction mechanisms of NORR toward N$_2$O and N$_2$ are illustrated in Figure 3b. For both byproducts, the precondition is the stable adsorption of NO-dimer on the active centers.[20,50] We therefore first study the adsorption free energy of NO-dimer by considering different adsorption configurations. As shown in Figure S7a, the configuration with both NO featuring N-end pattern is found to be most stable for all the systems. For Ga-DV-N$_2$C$_2$$^a$ and Ga-PP-C$_4$, the positive adsorption free energy of NO-dimer prevents the formation of byproducts and suggest their high selectivity for NORR toward NH$_3$. Except these two cases, other ten systems exhibit a negative adsorption free energy of NO-dimer, and we thus further study the subsequent hydrogenation steps of NORR toward N$_2$O and N$_2$ for the remaining ten SAC candidates following the reaction pathways shown Figure 3b. Figure S7b,c lists the $\Delta G_{max}$ of hydrogenation steps for NORR toward N$_2$O and N$_2$ via the corresponding most favorable reaction mechanisms. To show the selectivity of NORR directly, $\Delta G_{max-NH3}$ vs. $\Delta G_{max-N2O}$ and $\Delta G_{max-NH3}$ vs. $\Delta G_{max-N2}$ for the ten SAC candidates are plotted in **Figure 4**a,b. It can be seen that the systems shown in gray color in Figure 1e display $\Delta G_{max-NH3}$ > $\Delta G_{max-N2O}$/$\Delta G_{max-N2}$, rendering their poor selectivity of NORR toward NH$_3$ synthesis and thus their elimination. On the other hand, the systems shown in green color in Figure 1e exhibit $\Delta G_{max-NH3}$ < $\Delta G_{max-N2O}$/$\Delta G_{max-N2}$ presenting high selectivity for NORR toward NH$_3$ synthesis. In addition to byproducts N$_2$O and N$_2$ during NORR process, H$_2$ resulted from hydrogen evolution reaction (HER) is another byproduct that should be suppressed for improving the Faraday efficiency.[20,30] We therefore estimate the competition between NORR toward NH$_3$ and HER for these six candidates. Figure S8a shows the calculated $\Delta G_{*H}$ under the limiting potential of NORR toward NH$_3$ ($U_{L-NH3}$ = -$\Delta G_{max-NH3}$), and $\Delta G_{max-NH3}$ vs $\Delta G_{*H}$ is summarized in Figure S8b. HER process is significantly suppressed for all these six candidates, ensuring the excellent selectivity of NORR for NH$_3$. This combined with their high NORR activity makes these six systems (Figures 1c and 4c) very promising SACs for NORR toward NH$_3$. Their corresponding free energy diagrams and the optimized intermediates of these screened SACs during the NORR toward NH$_3$ are shown in Figure S9. We conclude that main group metal elements indeed can serve as compelling active centers of SACs for NORR, which is in sharp contrast to the common sense that nonmetal and transition-metal atoms are exclusively active centers for SACs. In addition, in contrast to previously reported metal bulk and TM-based electrocatalysts,[31-35] the presented candidate structures drive a barrier-free NORR process, and in contrast to the previous reported NORR electrocatalysts they avoid catalyst poisoning, as they show sizable $\Delta G_{*NO}$ (0...-1 eV compared to -2...-7 eV in competing systems[31,32,35,41]).

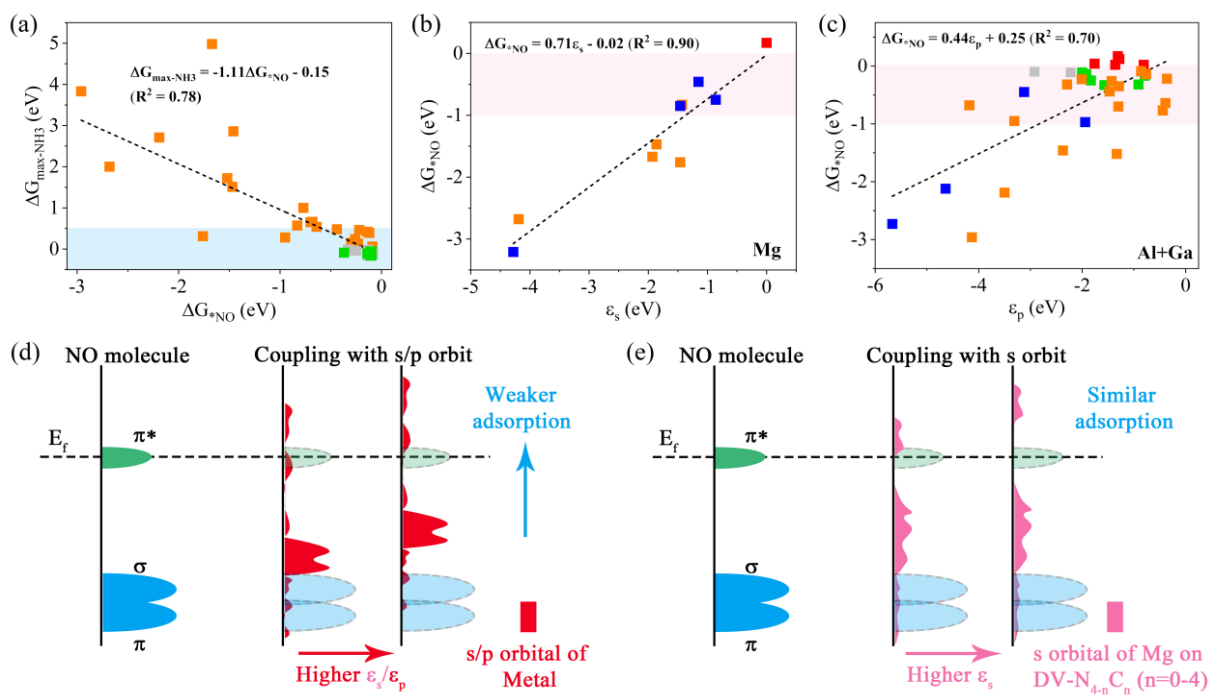

**Figure 5.** Descriptor and physical mechanism of catalytic activity. a) $\Delta G_{max-NH3}$ as a function of $\Delta G_{*NO}$. b) $\Delta G_{*NO}$ as a function of s-band centers of Mg on SV/PP-related SAC candidates. c) $\Delta G_{*NO}$ as a function of p-band centers of Al/Ga on SV/DV/PP-related SAC candidates. d),e) Schematic illustrations of the interactions between NO molecular orbitals and s/p-band of main-group metals. The color code of (a-c) is consistent with Figure 1d.

Figure S10 shows the electronic properties of these six SACs. While Al-DV-NC$_3$ is a semiconductor with indirect band gap of 0.7 eV, the other five systems are metallic with states of both spin polarizations crossing the fermi level. Therefore, all these six SACs harbor high conductivity, facilitating fast electron transfer during the reaction process.

The relationship between the $\Delta G_{*NO}$ and NORR activity are summarized in **Figure 5**a (for the candidates before the third screening step). It can be seen that there is a strong negative correlation between $\Delta G_{max-NH3}$ and $\Delta G_{*NO}$, suggesting $\Delta G_{*NO}$ as an excellent descriptor for NORR catalytic activity. It should be noted that for systems with high activity, $\Delta G_{*NO}$ lies within the range of 0…-1 eV. By examining the partial density of states (PDOS) of the s/p-band for main-group metals shown in Figure S11-S13, we find that $\Delta G_{*NO}$ is closely related to the s/p-band center ($\varepsilon_{s/p}$) of the main-group metal elements. The scaling relationships between $\varepsilon_{s/p}$ and $\Delta G_{*NO}$ for these 51 SAC candidates are illustrated in Figure S14. For the analysis of the valence electrons contributions we distinguish between the outer s for Mg and p electrons for Al/Ga atoms, with their respective orbital energies $\varepsilon_s$ and $\varepsilon_p$. Intriguingly, for Al/Ga-related systems, these is a strong linear relation between $\Delta G_{*NO}$ and $\varepsilon_p$ (Figure 5c). While for Mg-related systems, the relationship is a little complicated. In detail, $\Delta G_{*NO}$ is strongly correlated with $\varepsilon_s$ for Mg-SV-N$_{3-n}$C$_n$ and Mg-PP-N$_{4-n}$C$_n$ (Figure 5b), while it is rather poor for Mg-DV-N$_{4-n}$C$_n$ (Figure S14b). The same is true for a correlation between $\Delta G_{*NO}$ and $\varepsilon_p$ for the same systems (Figure S14f and S15).

The interaction between NO and the SAC's main-group metals reveals deeper insight into their catalytic activity. According to orbital analysis, the bonding state of NO adsorption on main-group metal center originates from the hybridization between the σ molecular orbital of NO and s/p-band of Mg/Al/Ga. Shapes and relative positions of NO molecular orbitals are slightly affected by the adsorption for most of the systems, and the σ orbital of NO mainly locates around -5…10 eV (Figures S16-S19). While the s/p-band contribution of main-group metals is almost negligible as it lies within the range of -5…-10 eV, the electrons of s/p-band of main-group metals, concentrated within 0…-5 eV, are responsible to the binding strength of NO, as illustrated in Figure S11-S13. Interestingly, we find that although the s/p-band of main group metals is generally delocalized,[51] the s/p-band of metals within the energy range of 0…-5 eV is relatively localized for Mg-SV-N$_{3-n}$C$_n$/-PP-N$_{4-n}$C$_n$ and Al/Ga-related systems. In this regard, as depicted in Figure 5d, by varying the s/p-band center ($\varepsilon_{s/p}$), the hybridization between the σ molecular orbital and s/p-band upon adsorption will be changed, thus affecting the interaction as well as the activation strength. In detail, with a high(low)-lying $\varepsilon_{s/p}$ modulated by less (more) band filling, the coupling strength is weakened (enhanced), yielding a low (high) activity for NORR toward NH$_3$. Therefore, the high NORR activity can be readily achieved through tuning the $\varepsilon_{s/p}$ in these systems. We wish to emphasize that the position of $\varepsilon_{s/p}$ can be effectively modulated in such systems through engineering of the N-coordination environment. On the other hand, $\varepsilon_{s/p}$ is an intrinsic physical parameter that is relevant to the NORR activity. Because s/p-band center can be directly measured in experiment, it should arise the attention for designing main-group metal elements as active centers for NORR.

In contrast to the other systems, for Mg-DV-$N_{4-n}C_n$, the delocalized character of the Mg s-bands is well preserved (Figure S11), and thus the coupling between NO electrons and s-band of Mg cannot be regulated by $\varepsilon_s$. Such feature generally is prone to broaden the adsorbate states,[22,24,25,52] yielding either too strong or too weak adsorption for adsorbates depending on the hybridization. But fortunately, the density of s-band is relatively low within the energy range of 0…-5 eV (Figure S11), thus the interaction between the Mg and adsorbed NO is moderate with $\Delta G_{*NO}$ between 0…-1 eV. With this result in hand, we understand why the $\varepsilon_s$ cannot define $\Delta G_{*NO}$ for Mg-DV-$N_{4-n}C_n$.

## 3. Conclusion

We comprehensively studied main-group metal elements as graphene-based active single atom catalyst centers for direct NO-to-$NH_3$ conversion. By means of first-principles calculations, out of 51 candidates we identified six SAC systems with excellent catalytic activity and selectivity for NORR. For these SACs, the NORR process proceeds spontaneously without any limiting potential, and with moderate $\Delta G_{*NO}$ of 0…-1 eV, which effectively avoids catalyst poisoning. Therefore, counter-intuitively, main-group metal elements indeed can serve as promising active centers for SACs. We rationalize the excellent performance of these systems to the modulation of s/p-band filling of the main-group metal centers by regulating coordination environment. Our rational four-step screening principle is generally applicable for exploring the possibility of introducing other main-group metals beyond Mg/Al/Ga to form SACs. We further discovered that the adsorption free energy of NO is an efficient catalytic descriptor for such SACs. The underlying physical mechanisms are revealed in detail. This work provides effective guidance for extending SACs to main-group metal elements as well as for designing high-efficient NORR catalysts.

## 4. Experimental Section

*Computational Methods:* Spin-polarized first-principles calculations are performed using the Vienna ab initio simulation package (VASP).[53] The projector augmented wave (PAW) method is used to describe the ion-electron interactions.[54,55] The exchange-correlation interactions are treated by the generalized gradient approximation (GGA) in form of Perdew-Burke-Ernzerhof (PBE) functional.[56] The plane-wave cutoff energy of 500 eV is adopted. Brillouin zone is sampled using a Monkhorst-Pack grid of 3 × 3 × 1. All structures are optimized with the convergence criteria of $10^{-5}$ eV and 0.01 eVÅ$^{-1}$ for energy and force, respectively. A vacuum space larger than 20 Å is employed to prevent adjacent interactions. Grimme's DFT-D3 correction is adopted to describe the van der Waals (vdW) interaction.[57] VASPKIT is adopted to obtain the free energy correction.[58] Details on the Gibbs free energy calculations can be found in the Supporting Information.

The s/p-band center of MGMs is defined as:

$$\varepsilon_{s/p} = \frac{\int_{-\infty}^{+\infty} E \times \rho_{s/p} dE}{\int_{-\infty}^{+\infty} \rho_{s/p} dE}$$

where $\rho_{s/p}$ is the density of s/p-band of main-group metals.

**Supporting Information**

Supporting Information is available from the Wiley Online Library or from the author.

**Acknowledgements**

This work is supported by the National Natural Science Foundation of China (Nos. 11804190, 12074217), Shandong Provincial Natural Science Foundation (Nos. ZR2019QA011 and ZR2019MEM013), Shandong Provincial Key Research and Development Program (Major Scientific and Technological Innovation Project) (No. 2019JZZY010302), Shandong Provincial Key Research and Development Program (No. 2019RKE27004), Shandong Provincial Science Foundation for Excellent Young Scholars (No. ZR2020YQ04), Qilu Young Scholar Program of Shandong University, and Taishan Scholar Program of Shandong Province. Deutsche Forschungsgemeinschaft is thanked for continuous support via CRC 1415.

**Conflict of Interest**

The authors declare no conflict of interest.

**Keywords**